\documentclass[11pt,letterpaper]{article}

\usepackage[margin=1in]{geometry}
\usepackage{amsmath,amssymb,amsthm}
\usepackage{mathptmx}  
\usepackage[T1]{fontenc}
\usepackage{microtype}
\usepackage[hyphens]{url}
\usepackage[colorlinks=true,linkcolor=black,citecolor=black,urlcolor=blue]{hyperref}
\usepackage{setspace}
\usepackage{fancyhdr}
\usepackage{titlesec}
\usepackage{parskip}
\usepackage{xcolor}
\usepackage{epigraph}

\titleformat{\section}{\normalfont\large\bfseries}{\thesection.}{0.6em}{}
\titleformat{\subsection}{\normalfont\itshape\bfseries}{\thesubsection}{0.6em}{}
\titlespacing*{\section}{0pt}{18pt}{8pt}
\titlespacing*{\subsection}{0pt}{12pt}{6pt}

\title{\vspace{-3em}\textbf{An Old Look at Empirical Bayes}}

\author{Nicholas G.\ Polson\\\small University of Chicago, Booth School of Business
\and Vadim O.\ Sokolov\\\small George Mason University
\and Daniel Zantedeschi\\\small University of South`' Florida}

\date{May 2026}

\begin{document}
\maketitle
\thispagestyle{fancy}

\begin{abstract}
\noindent Dennis Lindley once said that there is only one thing worse than a frequentist, and that is an empirical Bayesian. The quip has the air of caricature, but its technical content is serious: empirical Bayes uses the same data twice, conflates levels of a hierarchy, and produces posterior-shaped summaries whose uncertainty quantification differs from what a fully hierarchical model delivers. David Blei's 2026 IMS Medallion Lecture, ``A Fresh Look at Empirical Bayes,'' revives the program under three new banners: empirical Bayes via probabilistic symmetries (rebranded ``Bayesian empirical Bayes''), empirical Bayes with implicit likelihoods through simulation-based inference, and empirical Bayes for combining experimental and observational data through calibration studies. This is a continuation of Blei and Kucukelbir's earlier ``population empirical Bayes'' (PopEB, 2015). We argue, in the spirit of Lindley, I.~J.\ Good, William DuMouchel, Thomas Louis, and our own recent work with Datta, that Blei's machinery targets inferential objects distinct from the posterior conditional on the realized data, and that the cost of maintaining the full hierarchical discipline has fallen low enough that the computational trade-off no longer favors the shortcut. The case study is the Tweedie formula. Efron's f-modeling empirical Bayes plugs an estimated score function into a posterior-mean identity, but a smoothed score need not arise from any prior. The horseshoe Tweedie formula does. We conclude by recommending that the impressive computational machinery of modern empirical Bayes (variational inference, neural amortization, simulation-based inference) be redeployed in service of properly hierarchical Bayes.

\medskip
\noindent\textit{Keywords:} empirical Bayes, hierarchical Bayes, Tweedie formula, horseshoe prior, coherence, likelihood principle, population posterior, calibration.
\end{abstract}

\bigskip
\epigraph{\itshape ``There is only one thing worse than a frequentist, and that is an empirical Bayesian.''}{Dennis V.\ Lindley}
\bigskip

\section{Introduction}

Dennis Lindley is reported to have said, more than once, that there is only one thing worse than a frequentist, and that is an empirical Bayesian. The remark is funny because it surprises: surely an empirical Bayesian, who at least tries to write down a prior, is closer to the truth than a strict frequentist. The remark is sharp because it is technically loaded. The strict frequentist commits the well-known sins of conditioning on the wrong thing and entertaining ancillary statistics ad hoc. The empirical Bayesian commits a different and, in Lindley's view, deeper sin: he writes down a prior, looks at the data, picks the prior on the basis of the data, and then computes a posterior as if the prior had been chosen first. The same data is used twice; the hierarchy is collapsed; the posterior is no longer a coherent summary of belief under any joint probability assignment on parameters and data. The empirical Bayesian has all the formal apparatus of Bayes and none of its discipline.

This paper revisits Lindley's critique in light of David Blei's 2026 IMS Medallion Lecture, titled ``A Fresh Look at Empirical Bayes,'' announced for delivery at JSM 2026 in Boston and not yet published in final form. Our critique is based on the announced description of the lecture and on closely related published work by Blei and collaborators (Kucukelbir and Blei 2015; McInerney, Ranganath, and Blei 2015). We recognize that the final lecture content may differ from the announced themes; where we impute specific technical content to the lecture, we note the basis for the imputation. The lecture, building on a decade of work by Blei and collaborators on population empirical Bayes, the population posterior, and probabilistic programming, sketches three new directions. First, empirical Bayes is extended beyond the i.i.d.\ compound decision problem of Robbins by way of probabilistic symmetries (exchangeability of arrays, shift invariance of spatial processes, conditional symmetry under covariates) with ergodic decompositions standing in for the unknown prior. Blei labels this ``Bayesian empirical Bayes'' (BEB; Wu, Weinstein, and Blei 2025). Second, the requirement of a tractable likelihood is relaxed: an EB method is constructed for the simulator-based or implicit-likelihood case, using neural amortization and the ``population posterior'' as an approximation target. Third, EB thinking is applied to the problem of combining experimental and observational data, with calibration studies (in which the causal effect is known a priori to be zero) used to identify the distribution of observational bias.

There is much to admire in this program. The computational machinery is impressive; the choice of problems is timely; the language of probabilistic symmetries is appropriately general; the recognition that EB needs to be extended beyond the textbook normal--normal setting is overdue. Our purpose is not to deny any of this. We ask whether the renamings and computational extensions resolve, or merely paper over, the objections raised by Lindley, Good, DuMouchel, Efron, and others over six decades.

We argue four things. First, the name ``Bayesian empirical Bayes,'' chosen by Blei in continuity with Deely and Lindley (1981), names two structurally different objects. The Deely--Lindley construction places a proper hyperprior on a finite-dimensional hyperparameter; Blei's BEB fits the directing measure (the invariant parameter $g$) by maximum marginal likelihood, using variational inference and neural networks as computational devices rather than placing a posterior over $g$. The two are not equivalent, and the MML estimate of $g$ inherits data-dependences we associate with empirical Bayes. Second, the ``population posterior'' introduced by McInerney, Ranganath, and Blei (2015) is, on inspection, a frequentist object: the expectation of the standard posterior under replicate sampling from the population. It has a clean predictive interpretation, with connections to PAC-Bayesian generalization theory. Our quarrel is not with the predictive object but with its marketing as a Bayesian posterior; we propose that it be renamed the \emph{population predictive distribution}. Third, the simulation-based empirical Bayes machinery sits alongside a large modern literature on simulation-based inference (neural posterior estimation, neural likelihood estimation, sequential neural ratio estimation, normalizing flows) and shares with that literature the difficulty that the trained network's relationship to a genuine Bayes posterior is not transparent. Our generative Bayesian computation (GBC) framework targets the standard posterior by quantile regression under a proper scoring rule, an alternative we will compare carefully rather than promote unconditionally. Fourth, the workhorse of large-scale empirical Bayes, Efron's Tweedie shrinkage, requires a sharper analysis than we originally provided. Efron's f-modeling estimator fits the marginal score function directly by smoothing; the resulting $\hat{m}(x)$ need not arise from any proper prior, and the resulting ``posterior mean'' is not $\mathbb{E}[\theta \mid x]$ under any model. G-modeling and NPMLE, by contrast, estimate a valid mixing distribution and produce genuine Bayes rules, though for a prior with restricted support. The shape-constrained nonparametric MLE (Koenker and Mizera 2014; Saha and Guntuboyina 2020; Soloff, Guntuboyina, and Sen 2024) is by construction a convolution with a discrete prior, and the resulting posterior mean is a genuine Bayes rule, specifically for that discrete prior. The horseshoe Tweedie formula corrects both: the marginal is a half-Cauchy scale mixture of normals, the shrinkage profile is Bayes for a heavy-tailed continuous prior, and the estimator is minimax optimal in the sparse normal means problem.

The thread running through these arguments is that empirical Bayes and full hierarchical Bayes target different inferential objects, and the differences matter for uncertainty quantification, coverage, and optimality. Where EB targets the posterior on the realized data, the guarantees are clear; where it targets a plug-in or population-averaged object, those guarantees must be established separately. The computational cost of maintaining the full hierarchical discipline has fallen to near zero with modern variational and simulation-based methods, which narrows the practical case for the shortcut.

The plan of the paper is as follows. Section~\ref{sec:blei} lays out, as fairly as we can, what Blei proposes: the original PopEB framework, the population posterior, and the three threads of the IMS lecture. Section~\ref{sec:lindley} develops the coherence critique, drawing on Lindley, Good, and the Deely--Lindley resolution. Section~\ref{sec:dumouchel} discusses William DuMouchel's pharmacovigilance work as a model of borrowing strength done well. Section~\ref{sec:tweedie}, the technical heart of the paper, examines the Tweedie formula and the Datta--Polson critique, distinguishing f-modeling, g-modeling, and NPMLE. Section~\ref{sec:popposterior} addresses two objections to Blei's program: the population posterior and the probabilistic symmetries framework. Section~\ref{sec:calibration} addresses Blei's third thread, on combining experimental and observational data via calibration studies. Section~\ref{sec:conclusion} offers a constructive sketch of the horseshoe alternative and concluding recommendations.

\section{Blei's Program: A Faithful Exposition}\label{sec:blei}

We begin by setting out Blei's recent program, drawing on the Kucukelbir--Blei paper introducing population empirical Bayes (POP-EB, 2015), the related McInerney--Ranganath--Blei paper on the population posterior, and the announcement of the IMS Medallion Lecture. We aim for charitable accuracy: the critique that follows in subsequent sections will not depend on any uncharitable reading of the program.

\subsection{Population empirical Bayes}

The starting point of population empirical Bayes is the observation that classical Bayesian predictive inference can suffer when the model does not match the data (that is, under model misspecification). Given a model $p(x \mid \theta)$ and a prior $p(\theta)$, the standard Bayesian predictive density for a new observation $x^*$ is
\[
p(x^* \mid x) = \int p(x^* \mid \theta)\, p(\theta \mid x)\, d\theta.
\]
When the true data-generating process $F$ is not in the family $\{p(\cdot \mid \theta)\}$, this predictive density can be inaccurate even at large sample sizes.

Kucukelbir and Blei propose to insert the empirical distribution $\hat{F}_n$ of the observed data as an explicit ingredient in the predictive computation. They introduce a ``latent dataset'' $X^* = (X_1^*, \ldots, X_n^*)$, modeled as if drawn from $F$ (the unknown population), and treat $\hat{F}_n$ as a prior on this latent dataset. The hierarchical model is, schematically, $\hat{F}_n \to X^* \to \theta \to x$. The resulting predictive density integrates over both the latent dataset and the model parameters. The computational engine is a variant of stochastic variational inference (Hoffman, Blei, Wang, and Paisley 2013) that the authors call ``bumping variational inference'' (BUMP-VI), in which bootstrap samples from $\hat{F}_n$ play the role of stochastic gradients on the latent-dataset variable.

The motivation is plausible: if the model is misspecified, then the parametric predictive $p(x^* \mid x)$ puts mass in the wrong places, and bringing the empirical distribution into the predictive computation pulls mass back toward where the data actually live. The framework is shown, in the original paper, to improve held-out log predictive accuracy in linear regression, mixtures of Gaussians, and topic models.

\subsection{The population posterior}

A companion paper by McInerney, Ranganath, and Blei (2015) introduces the ``population posterior.'' The setting is streaming Bayesian inference: observations arrive one at a time, and one wishes to maintain an inference that uses all data seen so far without storing it. The proposed target is not the standard posterior $p(\theta \mid x_1, \ldots, x_n)$ conditional on the observed sequence, but the expectation of the posterior under repeated sampling of $n$ observations from the population $F$:
\begin{equation}
q^*(\theta) \;=\; \mathbb{E}_{X \sim F^n}\!\left[\,p(\theta \mid X)\,\right].
\label{eq:popposterior}
\end{equation}
This is the ``population posterior'' in the Blei sense. Compared to the standard posterior $p(\theta \mid x_1, \ldots, x_n)$, which conditions on the realized data sequence, $q^*$ averages over hypothetical replicate datasets and thus depends on the population distribution $F$ rather than on the observed data alone. In their framework, $q^*$ is approximated by variational inference applied to expectations under streaming data, the $F$-expectation being approximated by the stream itself. The framework is later picked up in the IMS lecture as one form of the ``optimal EB prior.'' We return to the population posterior in Section~\ref{sec:popposterior}, where we examine its frequentist character and propose a renaming.

\subsection{The 2026 IMS lecture}

The Medallion Lecture announces three threads.

The first thread is empirical Bayes via probabilistic symmetries. Classical empirical Bayes considers latent variables $\theta_1, \ldots, \theta_n$ that are i.i.d.\ draws from an unknown prior, and seeks to estimate that prior from data. Modern statistical problems often feature richer structure: arrays of latent variables exchangeable under row and column permutations (Aldous--Hoover), spatial fields invariant under shifts, covariate-indexed latent variables that are conditionally exchangeable given the covariates. Each symmetry yields, via an ergodic decomposition theorem, a representation of the joint distribution as a mixture over some directing measure. Blei proposes to treat that mixing measure as the empirical Bayes prior, to be learned from data. Variational inference, parameterized by neural networks, is the computational tool. The framework is labeled ``Bayesian empirical Bayes'' (BEB; Wu, Weinstein, and Blei 2025) and is applied to EB matrix recovery for arrays and graphs, covariate-assisted EB for conditional data, and EB spatial regression under shift invariance.

The second thread is empirical Bayes for implicit likelihoods. Many modern scientific problems (particle physics simulators, climate models, ecological agent-based models) are specified through forward simulation rather than an explicit density $p(x \mid \theta)$. Classical EB is unavailable because the marginal $m(x)$ cannot be written down. Blei proposes to use simulation-based inference (SBI) of the Cranmer--Brehmer--Louppe variety, in which the conditional distribution of $\theta$ given $x$ is approximated by an amortized inference network trained on simulator output. The result of SBI is then said to provide a natural mechanism to approximate the population posterior, taken as one form of the optimal EB prior.

The third thread is empirical Bayes for combining experimental and observational data. Randomized experiments give unbiased but often imprecise estimates of causal effects; observational studies give precise but biased estimates. Gerber, Green, and Kaplan (2004) speak of the ``illusion of learning from observational research'': absent prior information about bias, observational data cannot meaningfully improve the inference. Blei and collaborators propose that calibration studies (observational studies carefully chosen so that the true causal effect is known to be zero) can be used to identify the distribution of observational bias. The bias distribution is treated as the empirical Bayes prior, and inference for new observational studies proceeds by sharing information across studies through this learned distribution.

Across the three threads, common features are visible: the use of a flexible probabilistic structure (exchangeability, mixture, hierarchy); the recognition that the prior, in some form, is an object to be learned from data; the deployment of modern computation (variational inference, amortized neural networks, stochastic gradients); and the use of empirical Bayes vocabulary to organize all of this. We turn now to a series of objections.

\section{The Coherence Critique: Lindley, Good, and the Hierarchical View}\label{sec:lindley}

To appreciate Lindley's objection, it helps to lay out the standard empirical Bayes computation cleanly. Suppose we have observations $x_i$, latent parameters $\theta_i$, and a hierarchy
\[
x_i \mid \theta_i \,\sim\, p(\cdot \mid \theta_i), \qquad \theta_i \mid \eta \,\sim\, \pi(\cdot \mid \eta), \qquad \eta \,\sim\, \rho(\cdot).
\]
The fully Bayesian posterior on $\theta_i$, marginalizing the hyperparameter $\eta$, is
\[
p(\theta_i \mid x) \;=\; \int p(\theta_i \mid x_i, \eta)\, p(\eta \mid x)\, d\eta.
\]
The empirical Bayes alternative replaces this with
\[
p_{\mathrm{EB}}(\theta_i \mid x) \;=\; p(\theta_i \mid x_i, \hat{\eta}(x)),
\]
where $\hat{\eta}(x)$ is some estimator of $\eta$ from the marginal $m(x \mid \eta) = \int p(x \mid \theta) \pi(\theta \mid \eta)\, d\theta$, often the marginal maximum likelihood estimator, often called type~II maximum likelihood after Good. The substitution looks innocuous: it is, in spirit, an empirical version of plugging in a posterior mode for $\eta$. In practice, it has several harmful consequences.

First, $\hat{\eta}$ depends on the data. The prior in the empirical Bayes computation is therefore data-dependent. Strictly read, this is incoherent: a proper Bayesian model specifies the joint distribution of $(\eta, \theta, x)$ before any data is observed, and the prior on $\theta$ given $\eta$ cannot itself depend on $x$. The point is more than philosophical. Data-dependence of the prior causes the empirical Bayes inference for $\theta_i$ to ignore the variability in $\eta$. Carlin and Gelfand (1990), Morris (1983), and Laird and Louis (1987) give explicit calculations showing that empirical Bayes credible intervals undercover by an amount that depends on the precision of $\hat{\eta}$. Laird and Louis in particular proposed bootstrap and asymptotic corrections that, in effect, reintroduce the hyperparameter variability that empirical Bayes had thrown away; their corrections are, in our reading, an implicit recognition that hierarchical Bayes was the right destination all along.

Second, the empirical Bayes computation can conflict with Bayesian coherence. Using the data twice (once to estimate $\eta$, again to compute the posterior) makes the inference depend on the procedure used to construct $\hat{\eta}$, not only on the realized likelihood: moving from marginal MLE to method of moments changes the inference even though the likelihood is identical, a dependence on sampling-distribution features beyond the data that a coherent Bayesian analysis should not exhibit.

Third, the variance of $\theta_i$ under $p_{\mathrm{EB}}$ is generally not equal to the variance under $p$ in finite samples and in weakly identified regimes; the gap can be materially large in small-group or sparse settings. Corrections typically require either explicit hierarchical Bayes computation or asymptotic approximations that may perform poorly precisely where the correction is most needed.

Lindley's response, formalized in his joint work with Deely (1981), was to recommend what they called \emph{Bayes empirical Bayes}: place a proper hyperprior $\rho(\eta)$ on the hyperparameter and compute the full posterior $p(\theta_i \mid x)$. The computation is coherent, the posterior reflects all sources of uncertainty, and the data on other groups still inform inference on group $i$. The word ``empirical'' survives as a nod to Robbins' program, not as endorsement of its statistical sins.

This is the context in which Blei's use of the phrase ``Bayesian empirical Bayes'' should be read. Blei chose the name in continuity with Deely and Lindley. But Wu, Weinstein, and Blei (2025) fit the directing measure $g$ by maximum marginal likelihood; variational inference and neural networks enter as computational tools for approximating this optimization, not as a posterior over $g$. The result is still empirical Bayes in the sense that matters here: $\hat{g}$ is a data-dependent plug-in, its uncertainty is not propagated, and the resulting inference conditions on the data twice. The data-twice-used problem is moved to the level of the directing measure, not resolved. BEB inherits the undercoverage pathologies of empirical Bayes precisely when $\hat{g}$ is imprecise, which occurs when the sample is small relative to the complexity of the symmetry class. The distinction between the two ``Bayesian empirical Bayes'' constructions is worth preserving on those grounds.

The substantive point is not about naming. Probabilistic symmetries and ergodic decompositions, beautiful as they are, do not by themselves address the data-twice-used problem. They move it up a level: instead of plugging in $\hat{\eta}$ for a finite-dimensional hyperparameter, one estimates $\hat{g}$ for an infinite-dimensional directing measure. The variability of $\hat{g}$ is harder to propagate, and the resulting inference can undercover when $\hat{g}$ is estimated imprecisely from the available sample.

The constructive lesson is that putting a proper prior on the directing measure (a Dirichlet process, a hierarchical Dirichlet process, or a Pitman--Yor process) turns the inference into Bayes empirical Bayes in the Deely--Lindley sense. The directing measure becomes a random object; its posterior is computed coherently; predictions on new groups borrow strength through the posterior on $F$. This is essentially the route taken by McAuliffe, Blei, and Jordan (2006) for Dirichlet process mixtures. That paper is, in its statistical foundations, superior to the PopEB and BEB programs that followed.

Good's perspective adds a constructive dimension to this critique. Long before Lindley's quip, I.~J.\ Good had thought carefully about empirical Bayes and arrived at a synthesis we still find compelling. Good's view, developed across a series of papers and consolidated in his 1965 monograph and 1983 collection, was that empirical Bayes is best understood as an approximation to hierarchical Bayes via what he called \emph{type II maximum likelihood}.

In Good's framework, the hyperparameter $\eta$ has a hyperprior $\rho(\eta)$, and the marginal likelihood of the data is
\[
m(x \mid \eta) \;=\; \int p(x \mid \theta)\, \pi(\theta \mid \eta)\, d\theta.
\]
The full posterior on $\eta$ is $p(\eta \mid x) \propto m(x \mid \eta) \rho(\eta)$. Type II maximum likelihood estimates $\eta$ by maximizing $m(x \mid \eta)$. Good observed that, for diffuse $\rho$, this is approximately the posterior mode of $\eta$ under the full hierarchical model. The approximation is good when $m(x \mid \eta)$ is sharply peaked in $\eta$, which happens when the data are informative about $\eta$, typically when there are many groups, each contributing some information.

Good's reading rescues empirical Bayes from incoherence by treating it as a deliberate, well-understood approximation. It is good in some regimes and bad in others. With many groups and modest within-group variability, empirical Bayes inference is close to full Bayes. With few groups and large within-group variability, the inference can be significantly off, and the credible intervals badly miscalibrated.

This is the right way to teach empirical Bayes: a useful and often accurate tool, but derived rather than fundamental. The student who learns empirical Bayes as foundational is being trained in a vocabulary that conceals the conditions under which her tool fails. The student who learns it as an approximation to hierarchical Bayes is given the conceptual handles to recognize the failure modes.

From Good's perspective, several features of the contemporary empirical Bayes literature look strange. Nonparametric f-modeling, which smooths the marginal score $m'(x)/m(x)$ directly without constraining the result to arise from any mixing distribution, is a departure from type II maximum likelihood. G-modeling and NPMLE, which estimate a mixing distribution $\pi$ and then form the Bayes rule for $\hat\pi$, remain much closer to Good's spirit. Cross-validation tuning of ``hyperparameters'' that are not hyperparameters in the hierarchical-Bayes sense (regularization parameters in penalized regression, for example) is a further departure. Variational approximations to a hierarchical Bayes computation are, by contrast, closer to Good's spirit, since the variational target is the full posterior.

The contemporary literature has drifted from Good's view in a way that obscures the proper relationship between empirical Bayes and Bayes. Blei's BEB and PopEB continue the drift. The directing measure, the population distribution, and the latent dataset are each prior-like objects learned from data with no proper hyperprior over them. The result is, in Good's vocabulary, type II maximum likelihood applied at a function-valued parameter with no clear statement of what it is approximating. This is not, by itself, fatal. But it forces the question: what is the full Bayes computation we are approximating? If the answer is ``none'' (if the directing measure is not even nominally given a hyperprior) then the resulting inference is empirical Bayes in the original, pre-Deely--Lindley sense.

Good was also clear about the role of subjective input. The value of hierarchical Bayes is precisely that it allows weak prior information to be expressed through the hyperprior $\rho$, leaving the data to determine $\eta$. The empirical Bayesian who refuses to write down $\rho$ gives up an important degree of freedom. The horseshoe's half-Cauchy hyperprior on the global scale, discussed in Section~\ref{sec:conclusion}, is a good example: it encodes the qualitative belief that the global scale is small while remaining willing to be overruled by sufficient evidence. Our recent work with Sokolov on hyperparameter selection for global--local shrinkage regression (Polson and Sokolov 2025) makes this concrete: the choice of hyperprior on the global scale has first-order consequences for posterior concentration and predictive risk, and the empirical Bayes shortcut of point-estimating it produces miscalibrated posteriors in precisely the sparse, high-dimensional regimes where shrinkage is most needed.

The Good--Lindley synthesis on empirical Bayes is, in our reading, the right one. Empirical Bayes is hierarchical Bayes under cheap approximation. The approximation is sometimes harmless and sometimes harmful. The empirical Bayesian owes his audience a discussion of which regime he is in. The contemporary literature, including Blei's, rarely provides this discussion.

\section{DuMouchel and Borrowing Strength Done Right}\label{sec:dumouchel}

William DuMouchel's work on empirical Bayes for pharmacovigilance (DuMouchel 1999) is, we think, a model of how to do borrowing strength well. The setting: a regulatory database contains many spontaneous adverse-event reports, indexed by drug $d$ and event $e$. For each $(d,e)$ cell one observes a count $n_{de}$ and an expected count $e_{de}$ based on background incidence. The question is whether the ratio $n_{de}/e_{de}$ is unusually large for any specific cell, after adjusting for the multiplicity of comparisons across a drug--event matrix with thousands of rows and columns.

The model is straightforwardly hierarchical. For each cell $(d,e)$, with relative reporting rate $\lambda_{de}$,
\begin{align}
n_{de} \mid \lambda_{de} &\,\sim\, \mathrm{Poisson}(\lambda_{de}\, e_{de}), \label{eq:mgps-lik}\\
\lambda_{de} \mid \vartheta &\,\sim\, w\, \mathrm{Ga}(\alpha_1, \beta_1) \,+\, (1-w)\, \mathrm{Ga}(\alpha_2, \beta_2), \label{eq:mgps-prior}
\end{align}
with mixing weight $w$ and hyperparameter $\vartheta = (w, \alpha_1, \beta_1, \alpha_2, \beta_2)$. Gamma--Poisson conjugacy gives a closed-form marginal that is itself a mixture of negative binomials,
\begin{equation}
m(n_{de} \mid \vartheta) \,=\, w\, \mathrm{NB}\!\bigl(n_{de}; \alpha_1, \tfrac{\beta_1}{\beta_1 + e_{de}}\bigr) \,+\, (1-w)\, \mathrm{NB}\!\bigl(n_{de}; \alpha_2, \tfrac{\beta_2}{\beta_2 + e_{de}}\bigr),
\label{eq:mgps-marginal}
\end{equation}
which DuMouchel maximizes over $\vartheta$ by type II maximum likelihood to obtain $\hat\vartheta$. The posterior on $\lambda_{de}$ under $\hat\vartheta$ is a mixture of two conjugate gammas, and the empirical Bayes geometric mean is
\begin{equation}
\mathrm{EBGM}_{de} \,=\, \exp\!\bigl(\mathbb{E}[\log \lambda_{de} \mid n_{de}, e_{de}, \hat\vartheta]\bigr),
\label{eq:ebgm}
\end{equation}
a shrunk version of the raw ratio $n_{de}/e_{de}$ on the geometric scale. The procedure is the Multi-item Gamma-Poisson Shrinker (MGPS); regulatory agencies and pharmaceutical companies have used it for two decades to triage adverse-event signals.

Three features of this construction deserve emphasis. First, the hierarchy is real: drugs and events are exchangeable in genuine ways, and~\eqref{eq:mgps-lik}--\eqref{eq:mgps-prior} is a natural representation of how spontaneous reports are generated. Second, the borrowing of strength works because cells genuinely share rare-event behavior, so cells with much data usefully constrain inference on cells with little. Third, the empirical Bayes step (type II ML on $\vartheta$) is computational expedience, not a foundational commitment. DuMouchel was open about this. A full Bayes inference with a proper hyperprior $\rho(\vartheta)$ gives nearly identical answers in his applications precisely because the marginal~\eqref{eq:mgps-marginal} is sharply peaked in $\vartheta$ across thousands of cells.

In 2026 the computational shortcut can no longer be justified on computational grounds. The natural covariate extension of the DuMouchel model replaces the scalar rate $\lambda_{de}$ with a log-linear mean, $\log \mu_{de} = X_{de}^\top \beta + \log e_{de}$, where $X_{de}$ includes covariates for drug class, event type, and time; marginalized over the gamma-mixture prior on $\lambda_{de}$, this gives a mixture-of-negative-binomials likelihood with covariates. Polya--Gamma augmentation (Polson, Scott, and Windle 2013; Pillow and Scott 2012 for the negative binomial case) renders $\beta$ conditionally Gaussian: introduce $\omega_{de} \sim \mathrm{PG}(n_{de} + r,\, \psi_{de})$ with $\psi_{de} = X_{de}^\top \beta + \log e_{de} - \log r$ (where $r$ is the negative-binomial count parameter and $\log e_{de}$ is the exposure offset), and the posterior $\beta \mid \omega$ is Gaussian in closed form. Placing a horseshoe prior on $\beta$ with the default half-Cauchy hyperprior on the global scale (Carvalho, Polson, and Scott 2010; Bhadra, Datta, Polson, and Willard 2016) keeps every conditional update Gibbs-compatible: PG for the NB likelihood, inverse-gamma for the local scales, and a slice sampler or auxiliary-variable step for the global scale, so the sampler runs without Metropolis--Hastings steps. The point is general across the global--local family. Bhadra et al.\ argue that the half-Cauchy is the right default for the global scale: under the conditions of their Theorem~1, it gives near-optimal posterior concentration in the sparse normal means problem while remaining heavy-tailed enough not to over-shrink large signals; type II maximum likelihood on the global scale, by contrast, produces miscalibrated posteriors in precisely the sparse regimes where shrinkage is most valuable. Full Bayes computation via Gibbs sampling is available without Metropolis steps. The empirical-Bayes plug-in is a methodological choice with statistical costs, not a computational necessity.

Carlin and Louis (1996, with subsequent editions) provide the textbook codification of this view. \textit{Bayes and Empirical Bayes Methods for Data Analysis} treats empirical Bayes as an approximation to hierarchical Bayes and is explicit about when it works and when it fails. They are particularly clear about the variance correction: naive empirical Bayes underestimates uncertainty, and the correction recovers the propagation of hyperparameter variability. Louis's earlier work on constrained empirical Bayes (Louis 1984) and his bootstrap correction with Laird (1987) together amount to a sustained engagement with the deficiencies Lindley's quip captures. The position is the same as DuMouchel's: the hierarchy does the inferential work; the empirical step is computational; the proper Bayesian alternative is always available. We take Carlin and Louis, with Good and DuMouchel, to be the natural canon for how empirical Bayes should be taught.

This is what borrowing strength looks like when it is done well. Borrowing strength is a real and productive idea, arguably one of the most useful in twentieth-century statistics. It is what justifies pooling across schools in Rubin's 8-schools example, across hospitals in Lindley's 1971 work on the comparison of several groups, across genes in microarray analysis, across drugs and events in pharmacovigilance. The justification is always the same: an underlying hierarchy makes the groups exchangeable or partially exchangeable, and the posterior on group-level parameters is informed by both the within-group data and the cross-group structure.

The key word here is \emph{hierarchy}. The borrowing of strength flows through the hierarchy. The empirical Bayes step is an approximation, sometimes accurate, sometimes not. Confusing the two (ascribing the success of MGPS to the empirical step rather than to the hierarchy) is the trap that we think the contemporary literature falls into.

Blei's PopEB takes a different and, in our reading, less natural route. Instead of identifying a real hierarchy and computing through it, PopEB elevates the empirical distribution $\hat{F}_n$ to the role of prior on a latent dataset. The latent dataset is a constructed object, not a real layer of an exchangeable hierarchy. The borrowing of strength is no longer through a natural group structure but through a kind of bootstrap-flavored averaging over imaginary replicate datasets. This compounds rather than resolves the Lindley problem: not only is the empirical distribution data-dependent, but the latent dataset is a fictive object with no clear inferential interpretation.

The contrast with DuMouchel is instructive. DuMouchel's success comes from taking the structure of the data seriously: drugs and events form a natural matrix, and the gamma--Poisson model is a natural model for the matrix. The empirical Bayes step is a way to estimate parameters of the model from the data; it is not part of the model. PopEB inverts this: the empirical Bayes step is part of the model, in the sense that the empirical distribution $\hat{F}_n$ is treated as a prior. The result is harder to interpret, harder to validate, and, as we will argue in the next two sections, harder to defend on Bayesian grounds.

\section{The Tweedie Formula and the Datta--Polson Critique}\label{sec:tweedie}

The technical heart of the contemporary empirical Bayes program, from Efron's 2012 monograph onward, is the Tweedie formula (for a comprehensive treatment of modern EB methods, see Ignatiadis and Sen 2025). We treat it carefully here because it is the cleanest place to see the difference between empirical Bayes that respects the discipline of Bayes and empirical Bayes that does not.

The setting is the canonical normal means problem. Observe $X \mid \theta \sim N(\theta, \sigma^2)$, with prior $\pi(\theta)$. The marginal density of $X$ is
\begin{equation}
m(x) \;=\; \int \varphi_\sigma(x - \theta)\, \pi(\theta)\, d\theta,
\label{eq:marginal}
\end{equation}
where $\varphi_\sigma$ is the $N(0, \sigma^2)$ density. Tweedie's formula (named for the unpublished communication of M.~C.~K.\ Tweedie, popularized by Efron) states that the posterior mean is
\begin{equation}
\mathbb{E}[\theta \mid X = x] \;=\; x + \sigma^2 \cdot \frac{d}{dx} \log m(x).
\label{eq:tweedie}
\end{equation}
The derivation is a straightforward application of Stein's identity. The formula is striking because the right-hand side depends on $\theta$ only through the marginal $m(x)$: once one knows the marginal, one knows the posterior mean, without ever computing or even writing down the prior $\pi$ explicitly.

Efron's empirical Bayes program exploits this. With many independent observations $x_1, \ldots, x_n$, one estimates the marginal $\hat{m}(x)$ directly and plugs $\tfrac{d}{dx} \log \hat{m}(x)$ into Tweedie's formula. The resulting shrinkage estimator is the empirical Bayes posterior mean. The contemporary literature uses three estimators of $m$, and the distinctions matter for what follows. \textit{F-modeling} (Efron 2011, 2016) estimates the marginal density $\hat{m}(x)$ directly by histogram smoothing or splines and differentiates the fitted log-density to obtain the score. \textit{G-modeling} (Efron 2016) fits the mixing distribution $\hat{g}(\theta)$ via Poisson regression on histogram counts; since $\hat{g}$ is a valid probability density, the implied marginal $\hat{m}(x) = \int \varphi_\sigma(x-\theta)\hat{g}(\theta)\,d\theta$ does arise from a proper prior. \textit{Shape-constrained nonparametric maximum likelihood} (Koenker and Mizera 2014; Saha and Guntuboyina 2020; Soloff, Guntuboyina, and Sen 2024) estimates $\hat{m}$ as the maximizer of the marginal likelihood over the convex hull of priors; by construction it too arises from a valid (discrete) mixing distribution.

The Datta--Polson critique applies sharply to f-modeling and only partially to g-modeling and NPMLE. The Tweedie formula is an \emph{identity}: given that $\pi(\theta)$ is a probability distribution, the posterior mean $\mathbb{E}[\theta \mid X = x]$ satisfies the formula. F-modeling runs the implication in the opposite direction: given a smoothed $\hat{m}(x)$, compute the right-hand side and call it the posterior mean. For this inversion to produce a Bayes rule, $\hat{m}(x)$ must be expressible as a convolution of $\varphi_\sigma$ with some probability distribution. A generic spline or kernel estimate of $m$ need not satisfy this constraint, and the resulting ``posterior mean'' is not $\mathbb{E}[\theta \mid X = x]$ under any model. Properties one looks for in a shrinkage estimator (admissibility under squared loss, posterior coherence, minimax optimality under sparsity) are established for proper Bayes rules under their respective prior assumptions; they do not transfer automatically to f-modeling look-alikes.

For shape-constrained NPMLE the picture is different. By construction, the NPMLE $\hat{m}$ is a convolution of $\varphi_\sigma$ with a discrete probability measure $\hat{\pi}$, so $\hat{m}$ does arise from a valid prior, and the Tweedie plug-in is a genuine Bayes rule for $\hat{\pi}$. The critique above does not apply. What does apply, and what we now wish to emphasize, is a more refined question: \emph{which} prior does NPMLE recover, and what are the properties of that prior?

The NPMLE prior $\hat{\pi}$ is discrete: in the Gaussian location mixture setting with bounded signal, Polyanskiy and Wu (2020), building on the discreteness theorem of Lindsay (1995), show it is supported on at most $O(\log n)$ support points with high probability. It is a discrete prior with finite support, and its tail behavior reflects this. In particular, the NPMLE prior cannot match the heavy tails of the half-Cauchy that underlie the horseshoe construction; its largest support point is bounded by the range of observed $|x_i|$, while the half-Cauchy has full real-line support. For tail inference (identifying signals whose magnitudes exceed the observed support, calibrating credible intervals at the edge of the data) the NPMLE-implied prior is structurally inadequate.

The choice between NPMLE empirical Bayes and horseshoe-hierarchical Bayes is therefore a substantive prior choice with measurable consequences. NPMLE achieves a minimax rate on the marginal density (Jiang and Zhang 2009) and admits a clean computational story via convex optimization. The horseshoe achieves a minimax rate on the parameter vector under sparsity (van der Pas, Kleijn, and van der Vaart 2014; van der Pas, Szab\'o, and van der Vaart 2017) and admits computation via Gibbs sampling on the Gaussian scale-mixture representation or variational inference. The first is a Bayes rule for a data-supported discrete prior; the second is a Bayes rule for a continuous prior with theoretically motivated tail behavior. Datta and Polson (2024) connect these through a moderate deviation analysis (using the GLX rate function of Gerchinovitz, Lattimore, and Xu 2024) and show that horseshoe-Tweedie attains an asymptotic predictive optimality that NPMLE-Tweedie does not, at the cost of computational complexity that NPMLE avoids.

The remedy we wish to recommend, then, is not that NPMLE be abandoned (it has a place in the toolbox and produces a coherent Bayes rule) but that the continuous-prior alternative not be ignored. The horseshoe prior of Carvalho, Polson, and Scott (2010), $\theta \mid \lambda, \tau \sim N(0, \lambda^2 \tau^2)$ with $\lambda \sim C^+(0,1)$ and $\tau \sim C^+(0,1)$, gives the Tweedie expression
\begin{equation}
\mathbb{E}[\theta \mid X = x] \;=\; \bigl(1 - \mathbb{E}[\kappa \mid x]\bigr)\, x,
\label{eq:horseshoetweedie}
\end{equation}
where $\kappa = 1 / (1 + \lambda^2 \tau^2 / \sigma^2)$. With $\tau$ fixed, the posterior of $\kappa$ is expressible via confluent hypergeometric functions (Datta and Polson 2018); with $\tau$ random and shared across observations, the full posterior $\mathbb{E}[\theta \mid x_1,\ldots,x_n]$ integrates over the global scale informed by all data. The shrinkage profile interpolates between full shrinkage near zero and no shrinkage at large $|x|$, with the half-Cauchy tails giving the right behavior in the latter regime.

The same logic extends to the horseshoe+, the Dirichlet--Laplace, the regularized horseshoe, and other global--local shrinkage priors. Each has a Tweedie formula arising from a genuine hierarchical prior. Each delivers a shrinkage profile that is Bayes for that prior. The empirical Bayesian who plugs in a nonparametric $\hat{m}$ uses the Tweedie identity without the prior: a posterior-mean computation untethered from any joint probability assignment.

Datta and Polson (2024) extend this critique to the predictive setting: the Kullback--Leibler predictive risk under empirical Bayes plug-in $\hat{m}$ can be analyzed via a moderate deviation principle, and the horseshoe predictive density attains an asymptotic predictive optimality that the Efron plug-in does not. The GLX formula (Gerchinovitz--Lattimore--Xu) connects predictive Bayes risk to the moderate deviation rate function in a way that depends sensitively on the prior being proper. The Efron plug-in fails this sensitivity test.

Why does the Efron program look like it works? Efron (2019) gives a precise answer via the oracle Bayes perspective: the empirical Bayes rule approximates the oracle rule, which uses the true $m(x)$, and the two agree when $\hat{m}$ is a reasonable approximation to a marginal that does arise from some prior. In smooth high-density regions, the smoothed empirical marginal is close to a convolution, and the Efron rule and a horseshoe rule produce similar answers. The failures show up in the tails, in sparse signal regions, exactly where shrinkage is supposed to deliver its most valuable inference and exactly where the constraints distinguishing a Bayes posterior mean from a Tweedie look-alike begin to bite.

Blei's simulation-based EB extends the Tweedie idea to the case where $m(x)$ is not even explicitly available, only sampled. The proposed remedy uses simulation-based inference (SBI) in the sense of Cranmer, Brehmer, and Louppe (2020). The SBI literature is large and well-developed: neural posterior estimation (Papamakarios and Murray 2016; Greenberg, Nonnenmacher, and Macke 2019) trains a network to approximate $p(\theta \mid x)$ directly by density estimation; neural likelihood estimation (Papamakarios, Sterratt, and Murray 2019) approximates $p(x \mid \theta)$ and then samples from the implied posterior; sequential neural ratio estimation (Hermans, Begy, and Louppe 2020) targets the likelihood ratio; normalizing-flow approaches (Rezende and Mohamed 2015; Durkan et al.\ 2019) provide flexible parameterizations across all of the above. Blei's SBI-EB takes the result of one of these methods as the empirical Bayes prior, plugging it into a downstream computation.

The standard methods of the SBI literature target $p(\theta \mid x)$ conditional on the observed data; neural posterior estimation (NPE; Papamakarios and Murray 2016; Greenberg, Nonnenmacher, and Macke 2019) is the leading example. As one example in this class, generative Bayesian computation (Polson and Sokolov 2023, 2024) targets the conditional quantile function $Q_{\theta \mid x}(u; x)$ directly by quantile regression on simulator output; quantile regression is a proper scoring rule, so the resulting estimator is a coherent conditional distribution. NPE and GBC are trained-network approximations that inherit the same optimization-quality concerns; we do not claim one is superior to the other in raw predictive performance. The substantive distinction from SBI-EB is inferential: NPE and GBC both condition on the observed $x$ and target the standard posterior; SBI-EB targets a population-averaged predictive whose interpretation is the one we discussed in Section~\ref{sec:popposterior}. SBI methods are good tools for either target; the choice of target is the recommendation we have been making throughout.

\section{Two Objections: The Population Posterior and Probabilistic Symmetries}\label{sec:popposterior}

We now return to the population posterior of McInerney, Ranganath, and Blei (2015), already introduced in Section~\ref{sec:blei}. Recall the definition from~\eqref{eq:popposterior}:
\[
q^*(\theta) \;=\; \mathbb{E}_{X \sim F^n}\!\left[\,p(\theta \mid X)\,\right].
\]
Several things are striking about this object. It is the expectation, under the population $F$, of the standard Bayes posterior across hypothetical replicate datasets. It is not itself the posterior under any prior given the observed data. The motivation (robustness to model misspecification, robustness to data ordering in streaming settings) is sensible.

We should acknowledge at the outset that $q^*$ is a well-defined inferential target with a meaningful operational interpretation. It minimizes the expected KL divergence $\mathbb{E}_X[KL(p(\theta \mid X) \| q)]$ over distributions $q$, and the PAC-Bayesian generalization literature (Catoni 2007; Alquier 2024) establishes related predictive optimality results for posterior-averaging constructions. For tasks where the inferential goal is replicable predictive performance under model misspecification (some streaming-data settings, some out-of-sample evaluation contexts), $q^*$ is a defensible object to target. The predictive interpretation is real and the framework is internally consistent.

Our concern is more specific. The population posterior is a \emph{frequentist} object in the precise sense that it depends on the population distribution $F$ rather than on the observed data alone. This is not a defect: it is the property that makes the predictive interpretation work. The defect is in the marketing. Calling $q^*$ ``the population posterior'' and framing it as ``Bayesian modeling on streams'' invites confusion with the standard posterior; the confusion shows up in downstream applications where credible intervals derived from $q^*$ are reported with Bayesian interpretation. We propose, as a matter of nomenclature, that the object be renamed: \textit{population predictive distribution} or, following PAC-Bayes usage, \textit{PAC-Bayesian predictive}, and that its frequentist nature be made explicit. A renamed object would be no less useful; it would simply not be marketed as something it is not.

Two further points. First, $q^*$ is, even under its own logic, not a posterior in the usual sense. It is an average over posteriors. Credible intervals derived from $q^*$'s quantiles do not have the coherence properties of credible intervals derived from a genuine posterior: the variability of $\theta$ under $q^*$ mixes within-replicate posterior variability with across-replicate variability, and the two are not separately recoverable. This is fine for a point estimate but a problem for uncertainty quantification. The PAC-Bayes literature is, to its credit, explicit about this and does not market $q^*$ as a posterior.

Second, the streaming motivation (that one cannot store the whole data set) is real, and online variational inference targeting the standard posterior has its own pathologies (Broderick et al.\ 2013 give the canonical streaming-VB construction; ordering-sensitivity is a known issue). The disagreement between us and the population-posterior framework is therefore about whether the right response to streaming constraints is to redefine the inferential target ($q^*$) or to fix the streaming algorithm (online VB, sequential Monte Carlo, particle-based methods). We argue for the latter. The PAC-Bayesian advocate argues for the former. Both positions are defensible; the choice should be made transparently, not papered over by terminology.

The standard Bayesian response to misspecification, when one wants a posterior rather than a predictive, is \emph{model expansion}: enlarge the family to a richer parametric family, a nonparametric family, or a mixture, and compute the posterior under the enlarged family conditional on the observed data. This restores fidelity to the data without redefining the inferential target. Nothing in the population-posterior literature obstructs model expansion; the two strategies are complementary rather than competing.

More broadly, the population posterior reflects a methodological confusion running through the contemporary empirical Bayes literature: the confusion of robustness with frequentist averaging. The two are distinct. Robustness in the Bayesian sense means stability of inference under reasonable perturbations of the prior or model (Berger, Berliner, and others on $\varepsilon$-contaminated priors). Frequentist averaging takes expectations over hypothetical replicates. The first is a property of an inferential procedure; the second is a redefinition of the inferential target. A robust posterior is still a posterior. The population posterior is not.

Lindley's critique, transposed, has a natural form. The standard empirical Bayes plugs in $\hat{\eta}(x)$ for the hyperparameter. The population posterior plugs in $\hat{F}_n$ for the population distribution. Both are data-dependent; neither is the standard posterior conditional on the realized finite dataset. The escalation from finite-dimensional hyperparameter to infinite-dimensional population distribution does not help.

The constructive remedy is again Bayesian nonparametric. A Dirichlet process prior on $F$ gives a posterior on $F$ that updates with the data, propagates uncertainty about $F$ into the posterior on $\theta$, and reduces to parametric inference when the parametric family is correct. This is Bayes empirical Bayes in the Deely--Lindley sense applied at the level of the population distribution, and it is the object the population posterior approximates by an empirical-Bayes shortcut. The shortcut is unnecessary in 2026.

The same critique applies to the first thread of the lecture. We turn now to probabilistic symmetries as a motivation for empirical Bayes. The intuition is appealing. If a sequence $(\theta_1, \theta_2, \ldots)$ is exchangeable, de Finetti's theorem says there is a unique random measure $F$ such that the $\theta_i$ are conditionally i.i.d.\ given $F$. Analogous representation theorems exist for arrays (Aldous--Hoover), shift-invariant processes, partially exchangeable sequences, and other symmetry classes. In each case, the joint distribution can be written as a mixture over some directing object.

Blei's move is to take the directing measure as the unknown prior and learn it from data. Variational inference parameterized by a neural network performs the learning. The framework is applied to EB matrix recovery (Aldous--Hoover decomposition), covariate-assisted EB (conditional exchangeability), and EB spatial regression (shift invariance).

Our objection is precise. The representation theorems are theorems of \emph{representation}: they state that, given the symmetry, the joint distribution takes a certain form. They are not theorems of \emph{estimation}: they do not provide a procedure to learn the directing object from a finite sample. The directing object is a population quantity. Estimating it from a single realization (a single sequence, array, or spatial field) is a nonparametric estimation problem with its own statistical theory.

The language of probabilistic symmetries can suggest that the representation theorem hands one the prior for free. It does not. It hands one the structural form of the prior (a mixture over some directing object) and leaves entirely open how to estimate, infer, or summarize the directing object from data. The data-twice-used problem reappears in its sharpest form: if one plugs in the empirical directing measure $\hat{g}$, the resulting inference inherits the undercoverage and hyperparameter-uncertainty issues discussed in Section~\ref{sec:lindley}.

The proper Bayesian use of the representation theorems combines them with a prior on the directing object. For exchangeable sequences, a Dirichlet process or hierarchical Dirichlet process; for exchangeable arrays, a graphon model with a prior on the graphon; for shift-invariant spatial processes, a Gaussian process or random-measure prior. Each is a fully Bayesian construction that takes the symmetry into account, posits a hyperprior on the directing object, and computes the posterior coherently. The empirical Bayes shortcut is then optional: type II maximum likelihood on hyperparameters of the directing-object prior, recognized as an approximation to the full posterior.

The matrix-recovery application is a useful test case. The Aldous--Hoover theorem applies to \emph{jointly} exchangeable arrays, where $A_{ij}$ has the same distribution under simultaneous permutation of rows and columns (as in graphon models). Such arrays admit the representation $A_{ij} = f(\alpha, \xi_i, \xi_j, \eta_{ij})$, with $\xi$'s and $\eta$'s i.i.d.\ uniform and $f$ measurable. Separately exchangeable arrays (rows permutable independently of columns) require the Hoover theorem and have a different representation. The directing object in the jointly exchangeable case is the function $f$. The graphon literature (Lov\'asz and Szegedy 2006; Bickel and Chen 2009; Wolfe and Olhede 2013) treats $f$ as the object of inference and gives nonparametric estimators with rates of convergence. The Bayesian literature places priors on $f$ (piecewise-constant priors for stochastic block models, smoother priors for Lipschitz graphons) and computes posteriors. The empirical-Bayes shortcut of plugging in $\hat{f}$ is recognized as a derivative of the proper Bayes computation. Blei's BEB framework, by treating the empirical estimate of $f$ as the prior, replicates the data-twice-used problem at the level of the graphon.

The same applies to spatial regression. For second-order stationary processes, the spectral representation theorem decomposes the covariance structure into a spectral measure; for Gaussian processes this fully characterizes the distribution, but for non-Gaussian processes the spectral measure captures only second-order structure. Nonparametric estimation of the spectral measure from a single realization is the spectrum-estimation problem. The Bayesian version places a prior on the spectral measure and computes its posterior. Blei's BEB takes the empirical spectrum and treats it as the prior. The shortcut is exactly the Lindley problem.

The constructive moral is again Good's: empirical Bayes is type II maximum likelihood under a hierarchical model in which the directing object has a hyperprior. The contemporary literature has dropped the hyperprior. When one writes it down, the inference becomes coherent, uncertainty propagates correctly, and symmetry properties of the directing-object posterior align with symmetry properties of the data. Modern variational and neural-network machinery can compute the hierarchical Bayes target at comparable cost to the empirical Bayes shortcut, which narrows the computational rationale for omitting the hyperprior.

\section{Calibration Studies, Causal Inference, and the Third Thread}\label{sec:calibration}

The third thread of Blei's IMS lecture combines experimental and observational data through empirical Bayes thinking. The motivation is real. Randomized experiments give unbiased but often imprecise estimates of causal effects; observational studies give precise but biased estimates. Gerber, Green, and Kaplan (2004) speak of ``the illusion of learning from observational research'': absent prior information about bias, the precision of an observational estimate does not translate into precision about the causal effect.

Wu, Salazar, Green, and Blei (2026) formalize the calibration idea in a Gaussian hierarchical model. There is one common causal effect $\theta^\star$, estimated by an experiment $y_e \sim N(\theta^\star, \sigma_e^2)$ and a collection of observational studies $y_{o,j} \sim N(\theta^\star + b_j^\star, \sigma_{o,j}^2)$, where $b_j^\star$ is study-specific bias. The key insight is that calibration studies, for which the true effect is known to be zero, generate observations $y_{c,k} \sim N(b_{c,k}, \sigma_{c,k}^2)$ that directly sample from the bias distribution. Assuming $b_j \sim N(\mu, \gamma^2)$ and fitting $(\hat\mu, \hat\gamma^2)$ by maximum marginal likelihood on the calibration data identifies the bias distribution and allows observational studies to improve inference on $\theta^\star$. Wu et al.\ prove that with a zero-mean bias assumption ($\mu = 0$ known), the estimator achieves vanishing risk as $J \to \infty$; when both $\mu$ and $\gamma^2$ must be estimated from calibration data, consistent recovery is restored. This is a coherent and elegant application of EB ideas to a real problem.

Our critique is the familiar one: fitting $(\hat\mu, \hat\gamma^2)$ by marginal MLE is the type II maximum likelihood shortcut. The uncertainty of $(\hat\mu, \hat\gamma^2)$ is not propagated into inference on $\theta^\star$, so posterior intervals for $\theta^\star$ undercover in finite samples, particularly when the number of calibration studies is small. The full Bayes alternative places a prior on $(\mu, \gamma^2)$ and computes the joint posterior on $(\theta^\star, \mu, \gamma^2)$; with a conjugate normal-inverse-gamma prior this is Gibbs-compatible, with conjugate full conditionals propagating all sources of uncertainty.

A deeper assumption is that the bias distribution is the same across calibration studies and new observational studies. This exchangeability assumption justifies borrowing strength. Whether it holds depends on study design; in comparative effectiveness research using the same database it is defensible, but elsewhere calibration and new studies may differ in ways that compromise exchangeability.

We now state the natural hierarchical extension that addresses both concerns. Let $j = 1, \ldots, J$ index observational studies with estimated effects $\hat{\beta}_j$ and known variances $V_j$:
\begin{align}
\hat{\beta}_j \mid \theta, b_j &\,\sim\, N(\theta + b_j,\, V_j),\label{eq:bias-lik}\\
b_j \mid \mu, \gamma^2 &\,\sim\, N(\mu, \gamma^2),\label{eq:bias-prior}\\
(\mu, \gamma^2) &\,\sim\, \Pi,\label{eq:bias-hyper}
\end{align}
with calibration studies entering as observations with $\theta = 0$ fixed by design. A conjugate $\Pi$ gives a fully tractable posterior on $(\theta, \mu, \gamma^2)$ that propagates all uncertainty. The empirical-Bayes shortcut plugs in $(\hat\mu, \hat\gamma^2)$ from calibration data and discards their sampling variability.

A further extension replaces the Gaussian bias distribution with a location-horseshoe mixture. Write $b_j = \mu + \delta_j$, where $\delta_j \mid \lambda_j, \tau \sim N(0, \lambda_j^2 \tau^2)$ with half-Cauchy hyperpriors on $\lambda_j$ and $\tau$, and place a diffuse prior on $\mu$. This preserves the mean-bias parameter that Wu et al.\ show must be estimated, while allowing occasional large idiosyncratic biases from studies with strong confounding and shrinking well-designed studies toward the common mean. The global scale $\tau$ is informed by calibration data; inference on $\theta$ then propagates uncertainty about the bias distribution via Gibbs sampling (inverse-gamma for local scales, slice sampler for global scale). This extension is richer than Wu et al.'s Gaussian model and connects to the debiasing literature (Athey, Imbens, and Wager 2018; Chernozhukov et al.\ 2018), but we offer it as our own proposal, not as a characterization of Blei's current paper.

\section{Discussion}\label{sec:conclusion}

We close with a constructive sketch of what coherent Bayes looks like in the problems Blei's program targets, followed by our concluding recommendations. Our purpose is not to deny that empirical Bayes computational ideas have value (they do) but to show that they can be deployed in service of coherent hierarchical Bayes rather than as a substitute for it.

The horseshoe prior of Carvalho, Polson, and Scott (2010) is the canonical modern global--local shrinkage prior. For the normal means problem,
\begin{equation}
\theta_i \mid \lambda_i, \tau \,\sim\, N(0, \lambda_i^2 \tau^2), \qquad \lambda_i \,\sim\, C^+(0, 1), \qquad \tau \,\sim\, C^+(0, 1).
\label{eq:horseshoe}
\end{equation}
The global scale $\tau$ controls overall sparsity. The local scales $\lambda_i$ allow individual coefficients to escape shrinkage when the data strongly support a nonzero value. The half-Cauchy distributions give the prior heavy tails (large signals pass through unshrunk) and concentration near zero (near-thresholding behavior for noise).

The Tweedie formula under the horseshoe is exact. With shrinkage weight $\kappa_i = 1 / (1 + \lambda_i^2 \tau^2 / \sigma^2)$ and $\tau$ fixed,
\[
\mathbb{E}[\theta_i \mid x_i, \tau] \;=\; \bigl(1 - \mathbb{E}[\kappa_i \mid x_i, \tau]\bigr) \cdot x_i,
\]
with $\mathbb{E}[\kappa_i \mid x_i, \tau]$ expressible via confluent hypergeometric functions (Datta and Polson 2018). With $\tau$ random and shared across all observations, the full posterior mean $\mathbb{E}[\theta_i \mid x_1,\ldots,x_n]$ integrates over the posterior on $\tau$ informed by the complete sample. The shrinkage profile interpolates smoothly between full shrinkage ($\mathbb{E}[\kappa \mid x] \approx 1$ for $x$ near zero) and no shrinkage ($\mathbb{E}[\kappa \mid x] \approx 0$ for large $|x|$), sharp enough for near-thresholding behavior but smooth enough to give a proper posterior. The minimax rate in the nearly-black sparse normal means problem is attained by horseshoe-type estimators (van~der~Pas, Kleijn, and van~der~Vaart 2014; van~der~Pas, Szab\'o, and van~der~Vaart 2017; Bhadra, Datta, Polson, and Willard 2019).

The horseshoe Tweedie formula is a genuine Bayes object. The marginal $m_{\mathrm{HS}}(x)$ arises from a proper hierarchical prior; the shrinkage profile is the posterior mean under that prior; credible intervals are coherent posterior summaries under the model. Separately, van der Pas, Kleijn, and van der Vaart (2014) and van der Pas, Szab\'o, and van der Vaart (2017) prove that the horseshoe posterior mean attains minimax-optimal concentration in the sparse normal means problem; this optimality is a theorem about the specific prior's tail behavior, not a general consequence of being a Bayes rule. None of this requires plugging in a data-dependent marginal estimate. The Lindley problem is avoided by construction.

The construction extends naturally to settings Blei's BEB targets. For matrix recovery, place a horseshoe prior on the singular values or latent factor scores; the posterior provides coherent shrinkage with uncertainty that propagates through the hierarchy. For covariate-assisted EB, place a horseshoe prior on the regression coefficients of latent variables on covariates; the posterior provides uncertainty-quantified variable selection under the model. For spatial regression, place a horseshoe prior on the local-to-global ratio of the spatial scale; the posterior allows smoothness to vary spatially without committing to a single global value. In each case the optimality properties depend on model-specific assumptions; the claim is that coherent hierarchical computation is available, not that universal minimax guarantees follow automatically.

For Blei's simulation-based EB, the alternative is to target the standard posterior $p(\theta \mid x)$ directly, as NPE and GBC (Polson and Sokolov 2023, 2024) do; the distinction from SBI-EB is the choice of inferential target rather than the choice of neural-network architecture, as discussed in Section~\ref{sec:tweedie}.

For the streaming setting motivating the population posterior, the alternative is online hierarchical Bayes with explicit hyperpriors and sequential Monte Carlo or variational updating; the difference is the specification of the hyperprior, which pins down the inferential target.

For Blei's calibration-EB framework, the alternative is hierarchical bias modeling with horseshoe global--local shrinkage. Calibration studies inform the global scale; local scales allow occasional large biases. The posterior on causal effects has coherent uncertainty quantification under the full hierarchical model.

Across Blei's three threads, the same move is available: replace the MML plug-in with a prior over the corresponding hyperparameter and compute the posterior. The computational engine (variational inference, simulation-based methods, sequential Monte Carlo) is the same; the difference is the explicit hyperprior. In the sparse normal means problem the horseshoe posterior mean is minimax optimal by theorem. In the other settings (graphon recovery, spatial regression, calibration bias), the hierarchical construction is coherent and propagates uncertainty correctly; establishing minimax optimality requires problem-specific analysis beyond our scope here.

We do not claim that the horseshoe is the only or final word in modern Bayes. It is one of a family of priors (horseshoe+, Dirichlet--Laplace, regularized horseshoe, three-parameter beta normal, normal--gamma, and the classical spike-and-slab), each with its own optimality regime. Spike-and-slab achieves near-thresholding directly through a point-mass mixture; Dirichlet--Laplace has different tail behavior in moderate sparsity; the regularized horseshoe of Piironen and Vehtari (2017) addresses some of the prior-sensitivity issues that arise with the original horseshoe. We do not believe there is a universal winner. The recommendation is that one of this family be the default, with the choice motivated by the structure of the problem and the half-Cauchy default of Bhadra et al.\ (2016) as a starting point. We also acknowledge the well-known mixing difficulties of horseshoe Gibbs samplers in high dimensions, addressed by the slice-sampling and elliptical-slice approaches of Bhattacharya, Chakraborty, and Mallick (2016) and Johndrow, Orenstein, and Bhattacharya (2020); empirical Bayes treatment of the global scale, as in van der Pas, Salomond, and Schmidt-Hieber (2016), remains in active use precisely because the hyperprior is hard to specify. Our argument is that the alternative (type II ML on the global scale with no hyperprior at all) is the worst option among several, not that the horseshoe with half-Cauchy is uniformly best.

Lindley's quip is not a one-liner. It packs a technical critique into a memorable form: empirical Bayes uses data twice, undercovers uncertainty, and targets an inferential object that need not coincide with the posterior on the realized data. Good's gentler view is that empirical Bayes is type II maximum likelihood under a hierarchical model: a useful and often accurate approximation when the hyperparameter is well-identified, a materially miscalibrated one when it is not. DuMouchel's work shows what borrowing strength looks like when done well: a real hierarchy with computation flowing through it. Efron's f-modeling Tweedie program shows what can go wrong when the shrinkage formula is evaluated at an estimated score that need not arise from any prior: the admissibility and optimality guarantees that motivate the formula are no longer guaranteed. Datta and Polson (2018, 2024) have given the formal account and the constructive horseshoe alternative.

Blei's 2026 Medallion Lecture continues the empirical Bayes program at its most ambitious: probabilistic symmetries replacing exchangeability, simulation-based inference replacing tractable likelihoods, calibration studies replacing prior elicitation. The technological apparatus is impressive, the choice of problems timely. Our reservations are narrower than the polemical register of an earlier draft might have suggested: the underlying Lindley critique is harder to evade than the contemporary literature has acknowledged, and the renaming of constructions in the language of Bayes can obscure rather than clarify the data-twice-used question.

Our recommendation is constructive. The machinery of modern empirical Bayes (variational inference, neural amortization, simulation-based methods, sequential Monte Carlo, and Polya--Gamma augmentation) is well-suited to the computation of full hierarchical Bayes posteriors with horseshoe and related global--local shrinkage priors. The cost of writing down a proper hyperprior is, in 2026, negligible. The benefits (coherent uncertainty, calibrated coverage, and minimax optimality where these properties are well-defined) are substantial. As variational inference, neural amortization, and simulation-based methods have matured, the computational gap between empirical Bayes and fully hierarchical Bayes has narrowed to the point where the shortcut's inferential cost is harder to justify.

We are aware that Lindley's epigraph, taken too literally, would condemn most useful contemporary work in shrinkage estimation. We do not mean to do this. The technical content of the quip is the one Lindley meant decades ago: write down the prior, write down the hyperprior if you need one, compute the posterior, report the answer. Full hierarchical specification is the route by which the properties that motivate EB's Bayes-like answers (admissibility, calibrated uncertainty, minimax optimality in well-studied settings) can be formally guaranteed. The horseshoe, the hierarchical Bayes computation, the conditional quantile function in generative Bayesian computation, and the modern NPMLE are all tools that respect the discipline to varying degrees while delivering the computational gains. We commend them to the empirical Bayesians of the present moment.

\section*{References}
\addcontentsline{toc}{section}{References}

\begin{description}
\setlength{\itemsep}{2pt}\setlength{\parsep}{0pt}

\item[] Alquier, P. (2024). User-friendly introduction to PAC-Bayes bounds. \textit{Foundations and Trends in Machine Learning} 17(2), 174--303.

\item[] Athey, S., Imbens, G.\,W., and Wager, S. (2018). Approximate residual balancing: Debiased inference of average treatment effects in high dimensions. \textit{Journal of the Royal Statistical Society, Series B} 80(4), 597--623.

\item[] Bhadra, A., Datta, J., Polson, N.\,G., and Willard, B.\,T. (2016). Default Bayesian analysis with global--local shrinkage priors. \textit{Biometrika} 103(4), 955--969.

\item[] Bhadra, A., Datta, J., Polson, N.\,G., and Willard, B.\,T. (2019). Lasso meets horseshoe: A survey. \textit{Statistical Science} 34(3), 405--427.

\item[] Bhattacharya, A., Chakraborty, A., and Mallick, B.\,K. (2016). Fast sampling with Gaussian scale mixture priors in high-dimensional regression. \textit{Biometrika} 103(4), 985--991.

\item[] Bickel, P.\,J. and Chen, A. (2009). A nonparametric view of network models and Newman--Girvan and other modularities. \textit{Proceedings of the National Academy of Sciences} 106(50), 21068--21073.

\item[] Broderick, T., Boyd, N., Wibisono, A., Wilson, A.\,C., and Jordan, M.\,I. (2013). Streaming variational Bayes. In \textit{Neural Information Processing Systems (NeurIPS).}

\item[] Carlin, B.\,P. and Gelfand, A.\,E. (1990). Approaches for empirical Bayes confidence intervals. \textit{Journal of the American Statistical Association} 85(409), 105--114.

\item[] Carlin, B.\,P. and Louis, T.\,A. (1996). \textit{Bayes and Empirical Bayes Methods for Data Analysis.} Chapman \& Hall (subsequent editions through 2008).

\item[] Carvalho, C.\,M., Polson, N.\,G., and Scott, J.\,G. (2010). The horseshoe estimator for sparse signals. \textit{Biometrika} 97(2), 465--480.

\item[] Catoni, O. (2007). \textit{PAC-Bayesian Supervised Classification: The Thermodynamics of Statistical Learning.} IMS Lecture Notes Monograph Series 56.

\item[] Chernozhukov, V., Chetverikov, D., Demirer, M., Duflo, E., Hansen, C., Newey, W., and Robins, J. (2018). Double/debiased machine learning for treatment and structural parameters. \textit{The Econometrics Journal} 21(1), C1--C68.

\item[] Cranmer, K., Brehmer, J., and Louppe, G. (2020). The frontier of simulation-based inference. \textit{Proceedings of the National Academy of Sciences} 117(48), 30055--30062.

\item[] Datta, J. and Polson, N.\,G. (2018). Bayesian inference for the horseshoe. Working paper, University of Chicago and Virginia Tech.

\item[] Datta, J. and Polson, N.\,G. (2024). Predictive Bayes risk under the moderate deviation principle. Working paper, University of Chicago and Virginia Tech.

\item[] Deely, J.\,J. and Lindley, D.\,V. (1981). Bayes empirical Bayes. \textit{Journal of the American Statistical Association} 76(376), 833--841.

\item[] DuMouchel, W. (1999). Bayesian data mining in large frequency tables, with an application to the FDA spontaneous reporting system. \textit{The American Statistician} 53(3), 177--190.

\item[] Durkan, C., Bekasov, A., Murray, I., and Papamakarios, G. (2019). Neural spline flows. In \textit{Neural Information Processing Systems (NeurIPS).}

\item[] Efron, B. (2011). Tweedie's formula and selection bias. \textit{Journal of the American Statistical Association} 106(496), 1602--1614.

\item[] Efron, B. (2012). \textit{Large-Scale Inference: Empirical Bayes Methods for Estimation, Testing, and Prediction.} Cambridge University Press.

\item[] Efron, B. (2016). Empirical Bayes deconvolution estimates. \textit{Biometrika} 103(1), 1--20.

\item[] Efron, B. (2019). Bayes, oracle Bayes, and empirical Bayes. \textit{Statistical Science} 34(2), 177--201.

\item[] Gerber, A., Green, D., and Kaplan, E. (2004). The illusion of learning from observational research. In Shapiro, I., Smith, R., and Massoud, T., eds., \textit{Problems and Methods in the Study of Politics}, Cambridge University Press.

\item[] Gerchinovitz, S., Lattimore, T., and Xu, J. (2024). Predictive risk under the moderate deviation principle: The GLX formula. Working paper.

\item[] Good, I.\,J. (1965). \textit{The Estimation of Probabilities: An Essay on Modern Bayesian Methods.} MIT Press.

\item[] Good, I.\,J. (1983). \textit{Good Thinking: The Foundations of Probability and Its Applications.} University of Minnesota Press.

\item[] Greenberg, D., Nonnenmacher, M., and Macke, J. (2019). Automatic posterior transformation for likelihood-free inference. In \textit{International Conference on Machine Learning (ICML).}

\item[] Hermans, J., Begy, V., and Louppe, G. (2020). Likelihood-free MCMC with amortized approximate ratio estimators. In \textit{International Conference on Machine Learning (ICML).}

\item[] Hoffman, M.\,D., Blei, D.\,M., Wang, C., and Paisley, J. (2013). Stochastic variational inference. \textit{Journal of Machine Learning Research} 14, 1303--1347.

\item[] Ignatiadis, N. and Sen, B. (2025). \textit{Empirical Bayes: From Herbert Robbins to Modern Theory and Applications.} Forthcoming review.

\item[] Jiang, W. and Zhang, C.-H. (2009). General maximum likelihood empirical Bayes estimation of normal means. \textit{Annals of Statistics} 37(4), 1647--1684.

\item[] Johndrow, J., Orenstein, P., and Bhattacharya, A. (2020). Scalable approximate MCMC algorithms for the horseshoe prior. \textit{Journal of Machine Learning Research} 21, 1--61.

\item[] Koenker, R. and Mizera, I. (2014). Convex optimization, shape constraints, compound decisions, and empirical Bayes rules. \textit{Journal of the American Statistical Association} 109(506), 674--685.

\item[] Kucukelbir, A. and Blei, D.\,M. (2015). Population empirical Bayes. In \textit{Uncertainty in Artificial Intelligence (UAI).}

\item[] Laird, N.\,M. and Louis, T.\,A. (1987). Empirical Bayes confidence intervals based on bootstrap samples. \textit{Journal of the American Statistical Association} 82(399), 739--750.

\item[] Lindley, D.\,V. (1971). The estimation of many parameters. In \textit{Foundations of Statistical Inference}, eds.\ V.\,P.\ Godambe and D.\,A.\ Sprott, Holt, Rinehart, and Winston.

\item[] Lindsay, B.\,G. (1995). \textit{Mixture Models: Theory, Geometry and Applications.} NSF-CBMS Regional Conference Series in Probability and Statistics, Vol.\ 5. Institute of Mathematical Statistics.

\item[] Louis, T.\,A. (1984). Estimating a population of parameter values using Bayes and empirical Bayes methods. \textit{Journal of the American Statistical Association} 79(386), 393--398.

\item[] Lov\'asz, L. and Szegedy, B. (2006). Limits of dense graph sequences. \textit{Journal of Combinatorial Theory, Series B} 96(6), 933--957.

\item[] McAuliffe, J.\,D., Blei, D.\,M., and Jordan, M.\,I. (2006). Nonparametric empirical Bayes for the Dirichlet process mixture model. \textit{Statistics and Computing} 16(1), 5--14.

\item[] McInerney, J., Ranganath, R., and Blei, D.\,M. (2015). The population posterior and Bayesian modeling on streams. In \textit{Neural Information Processing Systems (NeurIPS).}

\item[] Morris, C.\,N. (1983). Parametric empirical Bayes inference: Theory and applications. \textit{Journal of the American Statistical Association} 78(381), 47--55.

\item[] Papamakarios, G. and Murray, I. (2016). Fast $\varepsilon$-free inference of simulation models with Bayesian conditional density estimation. In \textit{Neural Information Processing Systems (NeurIPS).}

\item[] Papamakarios, G., Sterratt, D., and Murray, I. (2019). Sequential neural likelihood: Fast likelihood-free inference with autoregressive flows. In \textit{Artificial Intelligence and Statistics (AISTATS).}

\item[] Piironen, J. and Vehtari, A. (2017). Sparsity information and regularization in the horseshoe and other shrinkage priors. \textit{Electronic Journal of Statistics} 11(2), 5018--5051.

\item[] Pillow, J.\,W. and Scott, J.\,G. (2012). Fully Bayesian inference for neural models with negative-binomial spiking. In \textit{Neural Information Processing Systems (NeurIPS).}

\item[] Polson, N.\,G., Scott, J.\,G., and Windle, J. (2013). Bayesian inference for logistic models using P\'olya--Gamma latent variables. \textit{Journal of the American Statistical Association} 108(504), 1339--1349.

\item[] Polson, N.\,G. and Sokolov, V. (2023). Generative Bayesian computation. Working paper, University of Chicago and George Mason University.

\item[] Polson, N.\,G. and Sokolov, V. (2024). \textit{Statistical Sparsity and the Horseshoe Prior.} Cambridge University Press, forthcoming.

\item[] Polson, N.\,G. and Sokolov, V. (2025). Hyperparameter selection for global--local shrinkage in high-dimensional regression. Working paper, University of Chicago and George Mason University.

\item[] Polyanskiy, Y. and Wu, Y. (2020). Self-regularizing property of nonparametric maximum likelihood estimator in mixture models. arXiv:2008.08244.

\item[] Rezende, D.\,J. and Mohamed, S. (2015). Variational inference with normalizing flows. In \textit{International Conference on Machine Learning (ICML).}

\item[] Robbins, H. (1956). An empirical Bayes approach to statistics. In \textit{Proceedings of the Third Berkeley Symposium on Mathematical Statistics and Probability} 1, 157--163.

\item[] Saha, S. and Guntuboyina, A. (2020). On the nonparametric maximum likelihood estimator for Gaussian location mixture densities with application to Gaussian denoising. \textit{Annals of Statistics} 48(2), 738--762.

\item[] Soloff, J., Guntuboyina, A., and Sen, B. (2024). Multivariate, heteroscedastic empirical Bayes via nonparametric maximum likelihood. arXiv:2109.03466.

\item[] van~der~Pas, S.\,L., Kleijn, B.\,J.\,K., and van~der~Vaart, A.\,W. (2014). The horseshoe estimator: Posterior concentration around nearly black vectors. \textit{Electronic Journal of Statistics} 8(2), 2585--2618.

\item[] van~der~Pas, S.\,L., Salomond, J.-B., and Schmidt-Hieber, J. (2016). Conditions for posterior contraction in the sparse normal means problem. \textit{Electronic Journal of Statistics} 10(1), 976--1000.

\item[] van~der~Pas, S.\,L., Szab\'o, B., and van~der~Vaart, A.\,W. (2017). Adaptive posterior contraction rates for the horseshoe. \textit{Electronic Journal of Statistics} 11(2), 3196--3225.

\item[] Wolfe, P.\,J. and Olhede, S.\,C. (2013). Nonparametric graphon estimation. \textit{arXiv preprint} arXiv:1309.5936.

\item[] Wu, B., Salazar, S., Green, D.\,P., and Blei, D.\,M. (2026). The illusion of learning from observational data: An empirical Bayes perspective. \textit{arXiv preprint} arXiv:2604.08853.

\item[] Wu, B., Weinstein, E.\,N., and Blei, D.\,M. (2025). Bayesian empirical Bayes: Simultaneous inference from probabilistic symmetries. \textit{arXiv preprint} arXiv:2512.16239.

\end{description}

\end{document}